\documentclass{jfm}
\usepackage{color}
\usepackage{amstext}
\usepackage{amssymb}
\usepackage{graphicx}
\usepackage{amsmath}
\usepackage{pgfplots}
\usepackage{float}

\newcommand\const{\mathrm{const}}

\renewcommand{\theequation}{\arabic{section}.\arabic{equation}}

\begin{document}


{\title[Variations of Saffman's robot] {Variations of Saffman's robot: two mechanisms of locomotion?}}

\author[V. A. Vladimirov]{ V.\ns A.\ns V\ls l\ls a\ls d\ls i\ls m\ls i\ls r\ls o\ls v$^{1,2,3}$}

\affiliation{$^1$ Sultan Qaboos University, Oman,\ $^2$ University of York, UK,\ $^3$ University of Leeds, UK}

\date{August 23, 2019}

\setcounter{page}{1}\maketitle \thispagestyle{empty}

\begin{abstract}
We study two mechanisms of locomotion of a body in an inviscid fluid, which take place without the shedding of vorticity;
we consider two simple examples of robots which are able to move along a straight trajectory.
The first one consists of a sphere with an internal moving material point (actuator); this illustrates \emph{the recoil locomotion}.
The second robot represents a sphere moving along a thin rigid rod, this is aimed at illustrating \emph{the deformational locomotion}.
The latter appears since this motion of a sphere can be seen as a `soliton type' deformation, moving along the rod.
The equations of motion for both robots are the same, while the intervals of variables and parameters are different.
The first robot was introduced by Saffman, who also wrote that he was unable to find any exact solution for the deformational locomotion: our paper partially fills this gap.
Some previous papers emphasize that, in the cases of locomotion, the deformations must be not axisymmetric.
We consider only axisymmetric examples, which expands the range of the involved possibilities.
The simple construction of presented robots allows us to operate with the exact solutions only, which can play a `reference' role.
Our aim is to analyse the main notions and terminology, used in this high-impact research area.
One can see, how two considered types of locomotion can be linked to each other.
\end{abstract}

\section{Introduction. Recoil mechanism of locomotion.}

\begin{figure}
\centering
\includegraphics[scale=0.7]{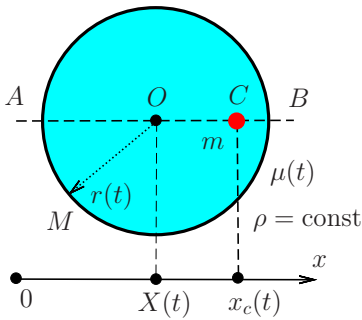}
\caption{Saffman's robot consists of a sphere of radius $r(t)$ and a movable material point (actuator) $C$ mounted inside the sphere.
Motions of points $O$ and $C$ take place along the line $AB$.
The distance $OC$ is $\xi(t)\equiv x_c(t)-X(t)$. The locomotion along the line $AB$ takes place due to the prescribed periodic oscillations of two control functions $r(t)$ and $\xi(t)$.}
\end{figure}
The locomotion of a body in an inviscid incompressible fluid, taking place without shedding vorticity, has a long history, see \emph{e.g.} \cite{Benjamin, Benjamin1, Saffman, Galper, Kelly, Kozlov, Chambrion, Vladimirov}.
In his pioneering paper, \cite{Saffman} studied two different cases of locomotion in an inviscid incompressible fluid.
His target was to show that the locomotion without shedding of vorticity is possible. In doing that, he considered separately the cases of heterogeneous bodies and homogeneous bodies.
In this paper we use a closely related, but different classification which separates two mechanisms of locomotion: \emph{the recoil locomotion} and \emph{the deformational locomotion}.
First, we consider \emph{Saffman's robot}, which represents the basic example of  recoil locomotion, and then we consider \emph{the spear robot}, which we use to demonstrate the deformational mechanism of locomotion.
We show that the identification of  the type of locomotion in particular cases can depend on the physical interpretation.
We discuss and analyse the involved notions and terminology.
All the papers \cite{Benjamin, Benjamin1, Saffman, Galper} emphasize that, in the cases of locomotion, the deformations must be not axisymmetric.
In this paper we consider only axisymmetric examples, which expands the range of the involved possibilities.
An advantage of the presented results is the use of only simple constructions of robots and only the exact solution of the equations of motion, which makes the mathematical results completely reliable.

\emph{Saffman's robot} consists of two parts: (i) a sphere of mass $M$, and (ii) an internal point mass (an actuator) of mass $m$, see Fig.1.
The space outside the sphere is filled with an inviscid incompressible fluid (of the density $\rho=\const$) quiescent at infinity;
the fluid flow is irrotational.
The sphere is deformable, such that it possesses radius $r(t)$, where $t$ is time. The virtual mass is $\mu(t)=\rho v(t)/2$ with $v(t)=4\pi r^3(t)/3$ (the validity of this expression for the time-dependent radius $r(t)$ is proven in \cite{Lamb, Saffman}).
For an additional physical purpose, \cite{Saffman} considered the one-dimensional locomotion of a three-dimensional ellipsoid with oscillating axes, but one which possessed a fixed volume.
This generalisation is not important for our consideration.
Let the motion of both the sphere and the actuator takes place along a fixed $x$-axis (along the diameter $AB$ in Fig.1), where $X(t)$ is an unknown coordinate of the sphere's center and $x_c(t)=X(t)+\xi(t)$ is a coordinate of the actuator.
Both functions $\mu(t)>0$ and $\xi(t)$ can be chosen arbitrarily (they are called control functions).
The physical restriction $\xi(t)<r(t)$ means that, at each instant $t$, the activator lays inside the sphere.
One may accept a stronger restriction $|\xi(t)|< \min r(t)$, while the exact location of the actuator inside the sphere does not play any role.
The conservation of linear momentum $P$ yields
\begin{eqnarray}\label{mom1}
P=(M+\mu(t))\dot X(t)+ m(\dot X(t)+\dot\xi(t))=0
\end{eqnarray}
where the dots above  letters stand for time-derivatives.
The total value of momentum $P$ is chosen to be zero.
This corresponds to the  motion, started from the state of rest of both robot and fluid.
Then:
\begin{eqnarray}\label{mom1a}
\dot X(t)=- \frac{m\dot\xi(t)}{M+m+\mu(t)}\equiv-\frac{\kappa\dot\xi(t)}{1+\lambda(t)}; \quad \kappa\equiv \frac{m}{M+m}, \ \lambda(t)\equiv\frac{\mu(t)}{M+m}
\end{eqnarray}
The right hand side of \eqref{mom1a} contains only known functions, so $X(t)$ can be calculated by direct integration.
Let both functions $\mu(t)$ and $\xi(t)$  be $T$-periodic.
With introduction of  frequency $\omega$ ($2\pi/\omega=T$) and variable $\tau\equiv \omega t$, this allows us to write
\begin{eqnarray}\label{mom1b}
X_\tau(\tau)=- \frac{m\xi_\tau(\tau)}{M+m+\mu(\tau)}\equiv-\frac{\kappa\xi_\tau(\tau)}{1+\lambda(\tau)}; \quad  \lambda(\tau)\equiv\frac{\mu(\tau)}{M+m}
\end{eqnarray}
where the subscript $\tau$ stands for $\tau$-derivative; $\xi$ and $X$ have the dimension of length.
Then, an averaging operation is defined as:
\begin{equation}\label{aver}
 \overline{f}\equiv \langle f\rangle\equiv (1/2\pi) \int_{\tau_0}^{\tau_0+2\pi} f(\tau) \ d\tau,\quad \forall \tau_0=\const
\end{equation}
The displacement $\Delta$ during one period is:
\begin{equation}\label{mom1c}
  \Delta\equiv X(\tau_0+2\pi)-X(\tau_0)=2\pi\langle X_\tau(\tau)\rangle =-2\pi\kappa\langle\xi_\tau(\tau)/[1+\lambda(\tau)]\rangle
\end{equation}
One can see that  if the control functions are mutually dependent,  then $\Delta\equiv 0$.
Hence for obtaining $\Delta\neq 0$ one should take them as mutually independent.
An example (similar to one presented in \cite{Saffman, Childress1}) is:
\begin{equation}\label{Ex2}
\xi=\xi_a\sin\tau, \quad \lambda=\overline{\lambda}+\lambda_a\cos\tau;\quad 0\le \lambda_a<\overline{\lambda}
\end{equation}
with constant amplitudes $\xi_a$ and $\lambda_a$, the later inequality has the physical meaning of non-negative virtual mass.
The evaluation of $\Delta$ \eqref{mom1c} yields
\begin{equation}\label{Ex2-1}
\Delta=\frac{2\pi \kappa\xi_a}{\lambda_a}\left(\frac{1}{\sqrt{1-\lambda_a^2/(1+{\overline{\lambda})^2}}}  -1 \right)
\end{equation}
which, for small amplitude $\lambda_a$, behaves as a linear function
\begin{equation}\label{Ex2-2}
\Delta\simeq\pi \kappa\xi_a\lambda_a/(1+\overline{\lambda})
\end{equation}
Then, as $\lambda_a$ increases, the displacement $\Delta$
monotonically increases until its natural maximin at $\lambda_a=\overline{\lambda}$.
The value of definite integral, leading to \eqref{Ex2-1}, follows from the formula 859.2 in \cite{Dwight}  and can also be calculated by \emph{Wolfram Mathematica 12}.
After $N$ periods, the robot can move itself to any required  (dimensionless) distance $N\Delta$.
One can notice that the dimensional distance, covered during a fixed time interval, is proportional to $\omega$.

\section{The spear robot: deformational locomotion versus recoil locomotion}

\begin{figure}
\centering
\includegraphics[scale=0.9]{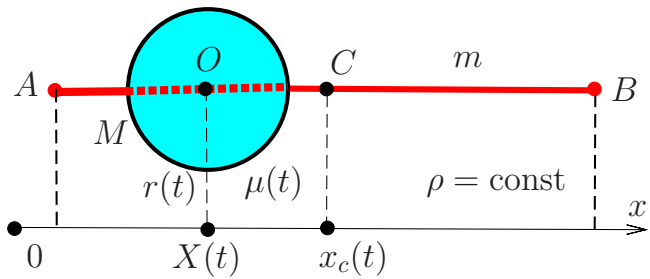}
\caption{\emph{The spear robot} consists of a sphere of variable radius $r(t)$ and a thin homogeneous rod $AB$ of fixed length $L$.
The distance between the center of gravity $C$ (with coordinate $x_c(t)$) of the rod and the center of the sphere $O$ (with coordinate $X(t)$) is a given function $\xi(t)\equiv x_c(t)-X(t)$.
The locomotion of the robot along the direction $AB$ is due to the prescribed changing of control functions $r(t)$ and $\xi(t)$.}
\end{figure}
\emph{The spear robot} consists of two elements: (i) a sphere of mass $M$ and a time-dependent radius $r(t)$, and (ii) a thin homogeneous rod of mass $m$ and fixed length $L$.
The rod penetrates through the axis (diameter) of the sphere, see Fig.2.
The  space outside the robot is filled with an inviscid incompressible fluid with the density $\rho=\const$; a fluid is quiescent at infinity, its motion is potential.
We consider a one-dimensional motion along the axis $x$ which is parallel to the rod.
Let $X(t)$ be the unknown coordinate of the sphere's centre and the position of the rod's center mass be $x_c(t)=X(t)+\xi(t)$.
We can choose two control functions $r(t)$ ($0\le r<\infty$) and $\xi(t)$ ($-L/2\le \xi\le L/2$) for obtaining the desired regime of robots' locomotion.
The conservation of momentum yields:
\begin{eqnarray}\label{mom2}
P=(M+\mu)\dot X+m(\dot X+\dot\xi)=0, \quad \mu=2\pi\rho r^3/3
\end{eqnarray}
where $\mu$ is  the virtual mass of the sphere.
The zero value of momentum corresponds to starting from a state of rest.
One can see that \eqref{mom2} and \eqref{mom1}  are the same.
This coinciding is not surprising, since the infinitely thin rod can be seen as an `external actuator', which doesn't interact with a fluid directly. Then we clarify the motion of the spear robot with four examples:
\begin{figure}
\centering
\includegraphics[scale=0.6]{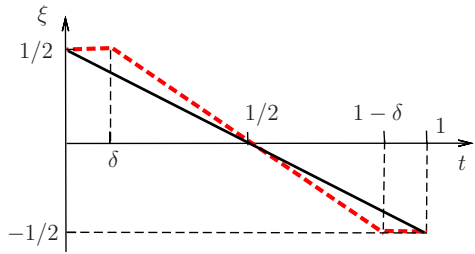}
\caption{Dependence $\xi(t)$: the continuous line corresponds to Example 2, the dashed line -- to  Example 3}
\end{figure}
\begin{figure}
\centering
\includegraphics[scale=0.6]{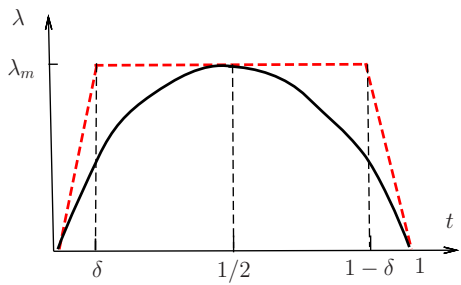}
\caption{Dependence $\lambda(t)$: the continuous line corresponds to Example 2, the dashed line -- to Example 3}
\end{figure}

\textbf{\emph{Example 1.}} \emph{Periodic functions $\lambda(t)$ and $\xi(t)$.}
Let the  oscillations of $\xi(t)$ and $\lambda(t)$ be $T$-periodic.
Then, the same equation \eqref{mom1b} for
the spear robot and for Saffman's robot leads us to the same solutions \eqref{mom1c}-\eqref{Ex2-2}, describing the recoil locomotion.
Hence, one can conclude that both robots can exhibit \emph{recoil locomotion}.
At the same time, this solution can be interpreted as the \emph{deformational locomotion} of the spear robot.
Indeed, it can be seen as a `soliton type' wave of the rod's deformation (having a spherical shape), having a time-dependent amplitude and oscillating along the rod.
Here one can see an interesting duality in the physical interpretations.

\textbf{\emph{Example 2.}} \emph{Parabolic $\lambda(t)$ and linear $\xi(t)$.}
At the same time, the spear robot possesses different modes of self-propulsion, which can be more certainly seen as \emph{deformational locomotion}.
For mathematical simplicity and for convenience in physical interpretation, we accept here that $M=0$ and hence will call the sphere \emph{a bubble}, like in \cite{Benjamin, Benjamin1, Galper}.
Let the bubble move from the left end of the rod ($\xi=L/2$) to the right end ($\xi=-L/2$) during time $T$, see Fig.3.
The dimensional variables are $X'=X/L, \xi'=\xi/L, t'=t/T$.
Then \eqref{mom1b} takes the form
\begin{eqnarray}\label{mom2b}
X_t(t)=-\xi_t(t)/(1+\lambda(t))
\end{eqnarray}\label{Vconst}
where all the `primes' are dropped for brevity and the subscript $t$ stands for derivative.
A simple translational motion of the bubble with constant speed $-\xi_t=1$
(in the dimensional variables the speed is $V=L/T$) is
\begin{eqnarray}\label{Vconst}
\xi(t)=1/2-t;\ \mbox{where}\  -1/2\le\xi\le 1/2,\ 0\le t\le 1
\end{eqnarray}
which means that the bubble moves with the constant velocity (equal unity).
The virtual mass is chosen as
\begin{eqnarray}\label{Vmu}
\lambda(t)=\lambda_m [1-(2t-1)^2]; \quad 0\le \lambda\le \lambda_m,\quad \lambda(0)=\lambda(1)=0
\end{eqnarray}
which means that the bubble is rising  from zero at $t=0$, has a maximum radius (corresponding $\lambda_m=\const$) at $t=1/2$, and then decreases to zero $r$ and $\lambda$ at $t=1$, see Fig.4.
Then, the integration of \eqref{mom2b} yields:
\begin{eqnarray}\label{XV}
&&\Delta(a)\equiv x_c(1)=\frac{1-a^2}{2a}\ln\frac{1+a}{1-a}-1,\quad \text{where}\ 0\le a^2\equiv \frac{\lambda_m}{(1+\lambda_m)}\le 1
\end{eqnarray}
The graph of $\Delta(a)$ is given in Fig 5.
As it should be physically, for $a\to 0$ (a very small bubble) the rod is not moving at all and $X(t)$ increases linearly from $-L/2$ to $L/2$.
For the opposite case, $a\to 1$ (a large bubble), it is not moving, and we have the rod shifted to the distance $-L$.
Finalizing this example, we write that during the period $T$, the center of mass of the spear robot is shifted to  distance $L\Delta(a)$.
Then, we can combine any number $N$ of such intervals, having the robot shifted to any (negative) distance $NL\Delta(a)$ with $a\equiv\sqrt{\lambda_m/(1+\lambda_m)}$.

\textbf{\emph{Example 3.}} \emph{Trapezoidal $\lambda(t)$ and piecewise $\xi(t)$.}
Here, we use the same dimensionless variables as in the previous example.
The time-dependence of the bubble's radius and the related virtual mass during the translational motion of the bubble may be confusing for some readers.
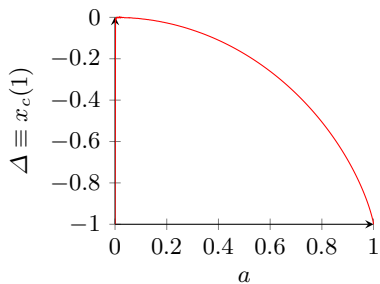
\begin{figure}
\begin{center}
\begin{tikzpicture}
\pgfplotsset{width=5cm,compat=1.9}
\begin{axis}[
    axis lines = left,
    xlabel = $a$,
    ylabel = {$\Delta \equiv x_c(1)$},
]
\addplot [
    domain=0:1,
    samples=500,
    color=red,
]
{(((1-x^2)/(2*x))*ln((1+x)/(1-x)))-1};
\end{axis}
\end{tikzpicture}
\end{center}
\caption{The dimensionless displacement of the center of gravity $x_c=X+\xi$ over one period as a function of parameter  $a^2=\lambda_m/(1+\lambda_m)$ in Example 2.}
\end{figure}
A simple way to avoid it is to choose:
\begin{eqnarray}
\lambda(t)=\left\{  \begin{array}{l}
\lambda_m t/\delta \\
\lambda_m \\
\lambda_m(1-t)/\delta \\
 \end{array}\right.,\
 \xi_t(t)=\left\{  \begin{array}{lcl}
0 & \mbox{in both cases}  & 0\le  t\le\delta  \\
-1/(1-2\delta) & \mbox{in both cases}  & \delta  <  t < 1-\delta  \\
0 & \mbox{in both cases}  & 1-\delta\le  t\le 1 \\
 \end{array}\right.
  \end{eqnarray}
for a real number $0<\delta<1$, see Figs 3 \& 4.
 This means that during the interval $0\le  t\le\delta$, the bubble stays at rest at $\xi=1/2$ and expands from zero to the size corresponding to $\lambda_m$.
During this procedure, the bubble generates a spherically symmetric point source type flow,  which doesn't affect the rod, hence the rod is not moving.
Then, the bubble moves with a constant velocity $-\xi_t=1/(1-2\delta)$ and constant virtual mass $\lambda=\lambda_m$ until it reaches the right end $\xi=-1/2$ at instant $t=1-\delta$.
Finally, at the right end of the rod the bubble stops (and hence the rod stops) and $\lambda(t)$ decreases to zero whilst generating the sink-type flow, which again does not affect the rod.
Then, the displacement of the rod during the time $T$ is $\Delta=-L/(1+\lambda_m)$.
This displacement $\Delta$ can be repeated any number of times $N$, which leads to the time-periodic locomotion of the spear robot and to the displacement $N\Delta$.
\begin{figure}
\centering
\includegraphics[scale=0.8]{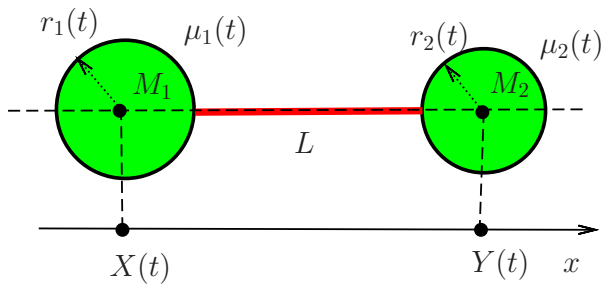}
\caption{\emph{The dumbbell robot} consists of two spheres of masses $M_1, M_2$ and radii $r_1(t), r_2(t)$ connected by a rigid rod of fixed length $L$. The control functions are $r_1(t)$, and  $r_2(t)$.}
\end{figure}

\section{Discussion}

Here we present one more example and few general comments.

\textbf{\emph{Example 4.}} \emph{The dumbbell robot.}
As a supplementary example of the deformational locomotion (however, leading only to an approximate solution) we mention here \emph{a dumbbell robot}.
It represents two spheres, connected by a thin weightless rod of constant length $L$, see Fig.6.
Taking the zero total momentum and the equality $Y-X=L+r_1+r_2$, one can derive an approximate equation
\begin{equation}\label{dumbbell}
  \dot X(t)=-\frac{(M_2+\mu_2(t))(\dot r_1(t)+\dot r_2(t))}{M_1+M_2 +\mu_1(t)+\mu_2(t)};\quad  \mu_i=2\pi r_i^3/3, \ i=1,2.
\end{equation}
obtained under the condition $L\gg\max(r_1,r_2)$ when the leading approximations for the virtual masses are given by that of a single sphere.
For the $T$-periodic control functions $r_1(t)$ and $r_2(t)$, the integration over $t$ immediately shows that such a \emph{dumbbell robot} demonstrates a locomotion when $r_1(t)$ and $r_2(t)$  represent independent functions. This robot gives us an instructive example of deformational locomotion.
The list of such examples can be continued, however they are out of the main scope of this paper, since even the virtual mass of a dumbbell is known only approximately (\cite{Lamb}).

Our general comments are:

1. In this paper, we present the conceptual examples, aimed at clarifying the general notions and terminology, such as those used in \cite{Benjamin, Saffman, Kelly, Galper}.
We have shown that in some cases the recoil locomotion and the deformational locomotion can be seen as different physical interpretations of the same motion.
For the full reliability of the results, we consider only exact solutions.

2. The \emph{spear robot} is capable of self-propulsion/locomotion in an inviscid incompressible fluid.
This robot can be seen as able to exhibit both recoil locomotion or deformational locomotion.
The latter appears since the motion of the bubble relative to the rod can be seen as a `soliton type' rod's deformations.
This deformation  has a spherical shape, and travels along the rod simultaneously changing its amplitude.
For the purpose of geometrical visualisation one may write that the time-evolution of the shape of spear robot looks somewhat similar to that in the self-propulsion of worms, see \emph{e.g.} \cite{Quillin, Box, Ziwang}.
Indeed, the worm's peristaltic propulsion is modelled by a wave, consisting of a sequence of radially expanded and radially contracted segments of its long cylindrical body.
In the spear robot, the motion of the sphere can be seen as a moving expanded segment of the rod.

3. \cite{Saffman} wrote that he was unable to build any exact solutions for deformational locomotion of a homogeneous robot.
Our paper partially fills this gap.
The spear robot cannot be homogeneous by its nature, however it can be seen as moving due to self-deformations.
 At the same time, the idea of a homogeneous body in the case of its time-dependent volume is physically not self-consistent, since it requires an instant redistribution of body's density in response to the changing volume.

4. There are several simplifications and restrictions accepted in our examples.
(i) The suggestion $M=0$ and the related use of the term \emph{bubble} do not represent any principal difficulty.
One can consider $M=\const$ or $M=M(t)$ as the result of mass exchange between the rod and the sphere.
(ii) The restriction of the time-dependent volume and virtual mass can be also avoided.
Indeed, one may consider, say, a thin disc that is perpendicular to the rod (instead of the sphere).

5. An interesting open problem is the generalization of the spear robot to creeping flows.
This problem is closely related to the classical examples of a self-propelling bending sheet, to peristaltic motion, to interaction between spheres \emph{etc.} see \cite{Childress, Childress2, Moffatt, MoffattNew, MofVlad}.
Those areas are well developed.
The analysis of related solutions for the general Navier-Stokes equations represents a great challenge, see \cite{Childress1}, where a similar geometry of the robot  (\emph{`a bug on a raft'}) is treated for a viscous fluid.

%

%

\begin{acknowledgments}
The author would like to express gratitude to
Profs. A.D.D. Craik, FRSE and
H.K. Moffatt, FRS, and Mr. A.A. Aldrick
for reading this manuscript and making useful comments.
Thanks to Profs. M. Al-Ajmi, I.A. Eltayeb, D.W. Hughes, K.I.Ilin, D. Kapanadze, and M.R.E. Proctor, FRS for helpful discussions.
This research is partially supported by the grant IG/SCI/DOMS/18/16 from SQU, Oman.
\end{acknowledgments}

\end{document}